\documentclass[%
reprint,
superscriptaddress,
nofootinbib,
amsmath,amssymb,
aps,
]{revtex4-2}

\usepackage{graphicx}
\usepackage{dcolumn}
\usepackage{bm}
\usepackage[caption=false]{subfig}
\usepackage{hyperref}
\usepackage{mhchem}
\usepackage{amsmath,amssymb}

\begin{document}

\title{Orbital-rotated Fermi-Hubbard model as a benchmarking problem for quantum chemistry with the exact solution}

\author{Ryota Kojima}
\author{Masahiko Kamoshita}
\author{Keita Kanno}
\email{kanno@qunasys.com}
\affiliation{QunaSys Inc.,
1-13-7, Hakusan, Bunkyo, Tokyo, 113-0001, Japan
}



\begin{abstract}
Quantum chemistry is a key target for quantum computing, but benchmarking quantum algorithms for large molecular systems remains challenging due to the lack of exactly solvable yet structurally realistic models. In particular, molecular Hamiltonians typically contain $O(N^4)$ Pauli terms, significantly increasing the cost of quantum simulations, while many exactly solvable models, such as the one-dimensional Fermi-Hubbard (1D FH) model, contain only $O(N)$ terms.
In this work, we introduce the orbital-rotated Fermi-Hubbard (ORFH) model as a scalable and exactly solvable benchmarking problem for quantum chemistry algorithms. Starting from the 1D FH model, we apply a spin-involved orbital rotation to construct a Hamiltonian that retains the exact ground-state energy but exhibits a Pauli term count scaling as $O(N^4)$, similar to real molecular systems.
We analyze the ORFH Hamiltonian from multiple perspectives, including operator norm and electronic correlation. We benchmark variational quantum eigensolver (VQE) optimizers and Pauli term grouping methods, and compare their performance with those for hydrogen chains. Furthermore, we show that the ORFH Hamiltonian increases the computational difficulty for classical methods such as the density matrix renormalization group (DMRG), offering a nontrivial benchmark beyond quantum algorithms.
Our results demonstrate that the ORFH model provides a versatile and scalable testbed for benchmarking quantum chemistry algorithms under realistic structural conditions, while maintaining exact solvability even at large system sizes.
\end{abstract}

\maketitle
\section{\label{sec:Intro}Introduction}

Quantum chemistry has emerged as a promising application area for quantum computing, with potential to outperform classical methods in simulating electronic structure problems. Recent years have seen rapid progress not only in quantum hardware, but also in algorithm development for both near-term noisy intermediate-scale quantum (NISQ) devices~\cite{Preskill2018quantumcomputingin} and future fault-tolerant quantum computers (FTQC). A wide range of quantum algorithms have been proposed for quantum chemistry, including those suited for FTQC~\cite{aspuru2005simulated, wang2008quantum, poulin2014trotter, babbush2015chemical} as well as NISQ-compatible approaches such as the variational quantum eigensolver (VQE)~\cite{peruzzo2014variational} and its variants~\cite{mcardle2020quantum, cerezo2021variational, bharti2022noisy, tilly2022variational} and also non-variational approaches such as the quantum-selected configuration interaction (QSCI)~\cite{kanno2023quantum} and its variants~\cite{nakagawa2024adapt,robledo2024chemistry}.

To assess and guide the development of such algorithms, benchmarking plays a critical role, and various proposals of methods to benchmark NISQ algorithms have been proposed~\cite{amico2023defining, lubinski2023application, lubinski2023optimization, Lubinski:2024hvc, dutt2023practical, wu2023variational, finzgar2202quark, gard2022classically, kobayashi2022parent}. In particular, evaluating algorithmic performance on electronic Hamiltonians of molecules or similar systems is crucial for applications in quantum chemistry. However, benchmarking in quantum chemistry presents a challenge: large-scale molecular systems of practical interest are beyond the reach of exact diagonalization, making it difficult to assess algorithmic accuracy without reliable reference solutions. Approximate methods such as coupled-cluster or density functional theory can be used as proxies, but introduce uncertainty and algorithm-dependent bias into the evaluation.

A natural solution is to use exactly solvable model Hamiltonians whose ground-state properties can be computed efficiently on classical computers. One well-known example is the one-dimensional Fermi-Hubbard (1D FH) model~\cite{hubbard1964electron, gutzwiller1963effect, kanamori1963electron, essler2005one}, which admits exact solutions via the Bethe ansatz~\cite{lieb1968absence, yang1967some}. However, the 1D FH model contains only $O(N)$ terms for an $N$-site system, whereas molecular electronic Hamiltonians generally contain $O(N^4)$ terms for systems with $N$ spatial orbitals, due to the four-index nature of the two-electron integrals. This large number of terms has practical consequences, such as increasing the number of measurements in NISQ algorithms~\cite{PhysRevResearch.4.033154} or the gate cost in FTQC algorithms~\cite{reiher2017elucidating}.

To overcome this discrepancy, we propose a model that retains the exact solvability of the 1D FH model while introducing a molecular-like $O(N^4)$ term structure. Our construction involves applying a spin-involved orbital rotation to the original 1D FH Hamiltonian, resulting in a new Hamiltonian, which we refer to as the orbital-rotated Fermi-Hubbard (ORFH) model, that preserves the ground-state energy but exhibits significantly more terms in its operator expansion. This enables us to bridge the gap between simplified solvable models and more realistic molecular Hamiltonians.

In this paper, we investigate the ORFH model as a surrogate benchmark problem for quantum chemistry. We study its similarity to molecular Hamiltonians in terms of operator norms, electronic correlation, and measurement cost. We also examine how the orbital rotation affects the computational difficulty for both quantum (VQE) and classical (DMRG) ground-state solvers. 

Our results demonstrate that The ORFH model captures essential structural features of molecular Hamiltonians while remaining exactly solvable, making it a scalable and controllable benchmark. It enables systematic evaluation of quantum and classical algorithms, such as analyzing how energy errors depend on bond dimension in DMRG or Trotter step size in quantum phase estimation.

The remainder of this paper is structured as follows. In Section~\ref{sec:Ham}, we introduce the ORFH Hamiltonian and explain its construction. Section~\ref{sec:Similar} investigates its structural and physical comparison to molecular Hamiltonians from various perspectives, including operator norm, electronic correlation, and benchmark of methods for grouping terms. In Section~\ref{sec:Numarical}, we benchmark both quantum (VQE) and classical (DMRG) algorithms using the ORFH Hamiltonian and compare their behavior with more traditional test cases such as hydrogen chains or the original FH model. We conclude with a summary and outlook in Section~\ref{sec:Concl}. Details of the numerical simulation settings are provided in the Appendix.

\section{\label{sec:Ham}Construction of the Orbital-Rotated Fermi-Hubbard Hamiltonian}

To design benchmark problems that closely mimic the complexity of molecular Hamiltonians, while still permitting exact classical solutions, we begin by revisiting the one-dimensional Fermi-Hubbard (1D FH) model. This model is not only exactly solvable via the Bethe ansatz but also captures essential features of strongly correlated electron systems. However, the original 1D FH Hamiltonian includes only $O(N)$ terms, which is far fewer than the $O(N^4)$ terms typically found in molecular Hamiltonians. In this section, we introduce a transformation, more precisely a spin-involved orbital rotation, that increases the number of Hamiltonian terms while preserving the ground-state energy, thereby enabling the construction of exactly solvable models with term structures comparable to those encountered in quantum chemistry.

\subsection{1D Fermi-Hubbard model}
We begin by introducing the one-dimensional Fermi-Hubbard (1D FH) model, which serves as the foundation for constructing our benchmark Hamiltonian. The Hamiltonian is given by
\begin{multline}
\label{eq:originalfh}
H = - t \sum_{i=0}^{N-1} \sum_{\sigma=\uparrow, \downarrow} (a^\dagger_{i, \sigma}  a_{i+1, \sigma} +  a^\dagger_{i+1, \sigma}  a_{i, \sigma})\\
- \mu \sum_{i=0}^{N-1} \sum_{\sigma=\uparrow, \downarrow}  a^\dagger_{i, \sigma} a_{i, \sigma} + U \sum_{i=0}^{N-1} a^\dagger_{i, \uparrow}  a_{i, \uparrow}  a^\dagger_{i, \downarrow} a_{i, \downarrow}
\end{multline}
where \( a^\dagger_{i, \sigma} \) and \( a_{i, \sigma} \) denote the fermionic creation and annihilation operators for site \( i \) and spin \( \sigma \), respectively. Here, \( N \) is the number of lattice sites, \( t \) is the hopping amplitude, \( \mu \) is the chemical potential, and \( U \) is the on-site Coulomb repulsion.
The symbols \( \uparrow \) and \( \downarrow \) denote spin-up and spin-down, respectively.
Throughout this work, we focus on the repulsive regime \( (U > 0) \), set \( t = 1 \), and assume periodic boundary conditions for simplicity.

While the Hamiltonian is expressed in terms of fermionic operators, it can be mapped to a qubit Hamiltonian using transformations such as the Jordan-Wigner mapping, yielding a system of \( 2N \) qubits.

In general, the Hilbert space dimension grows exponentially with system size, rendering exact diagonalization intractable for large \( N \). Nevertheless, the 1D FH model admits an exact solution via the Bethe ansatz~\cite{lieb1968absence, lieb2003one, yang1967some}, enabling efficient computation of the ground-state energy even for relatively large systems.

As discussed in the introduction, molecular Hamiltonians in quantum chemistry typically contain \( O(N^4) \) terms due to two-body interactions among \( N \) orbitals, and this term count significantly affects, e.g., the measurement cost in near-term quantum algorithms and the gate count in long-term ones. In contrast, the standard 1D FH model contains only \( O(N) \) terms, limiting its utility for benchmarking scenarios that aim to capture the measurement overhead of real molecular systems.

To bridge this gap, we introduce a spin-involved orbital rotation in the next subsection, which transforms the 1D FH Hamiltonian into a model with \( O(N^4) \) terms while preserving its exact solvability.

\subsection{Spin-Involved Orbital Rotation}
To increase the number of terms in the Fermi-Hubbard Hamiltonian while preserving its ground-state energy, we apply an orbital rotation involving spin degrees of freedom. This operation is implemented as a unitary transformation acting on the basis of fermionic creation and annihilation operators.

Since our orbital rotation involves both spatial and spin degrees of freedom, we relabel the operators as follows:
\begin{eqnarray}
c_{2i}^\dagger = a_{i, \uparrow}^\dagger, \quad     c_{2i + 1}^\dagger = a_{i, \downarrow}^\dagger, \\
c_{2i} = a_{i, \uparrow}, \quad     c_{2i + 1} = a_{i, \downarrow}.
\end{eqnarray}
Using a unitary matrix \( u \in \mathbb{C}^{2N \times 2N} \), the rotated operators are defined as
\begin{equation}
\label{eqs:orbital_rotation}
\tilde{c}_i^\dagger = \sum_{k=0}^{2N-1} u_{ik} c_k^\dagger, \quad
\tilde{c}_i = \sum_{k=0}^{2N-1} u_{ik}^* c_k,
\end{equation}
where the asterisk denotes complex conjugation.

It is important to note that spin-preserving orbital rotations do not increase the number of terms in the two-body part. In contrast, the spin-involved orbital rotation described here leads to a significant increase in the number of terms of $O(N^4)$. After applying the rotation, the Hamiltonian takes the form
\begin{equation}
\tilde{H} = \sum_{p,q=0}^{2N-1} h_{pq} c_p^\dagger c_q + \frac{1}{2} \sum_{p,q,r,s=0}^{2N-1} h_{pqrs} c_p^\dagger c_q c_r^\dagger c_s,
\end{equation}
where \( h_{pq} \) and \( h_{pqrs} \) are the one-body and two-body coefficients, respectively, given by
\begin{eqnarray}
h_{pq} &=& -\sum_{i=0}^{N-1} \Big(u_{2i,p} u_{2i+2,q}^* + u_{2i+2,p} u_{2i,q}^* \nonumber \\
&& \quad + u_{2i+1,p} u_{2i+3,q}^* + u_{2i+3,p} u_{2i+1,q}^* + \mu \delta_{pq} \Big), \\
h_{pqrs} &=& 2U \sum_{i=0}^{N-1} u_{2i,p} u_{2i,q}^* u_{2i+1,r} u_{2i+1,s}^*.
\end{eqnarray}
This expression has the same form as a general second-quantized molecular Hamiltonian used in quantum chemistry. Notably, the number of Pauli terms in the qubit representation of \( \tilde{H} \) scales as \( O(N^4) \), in contrast to the \( O(N) \) scaling of the original Fermi-Hubbard model. This increase enables benchmarking of quantum chemistry algorithms under conditions more closely aligned with molecular systems.

A key difference between the original and rotated Hamiltonians lies in spin symmetry. While the original Fermi-Hubbard Hamiltonian conserves the total spin-\( z \) component \( S_z \), the spin-involved orbital rotation generally breaks this symmetry, and \( S_z \) is no longer a good quantum number.

Since the rotation is implemented via a unitary transformation, the spectrum of the Hamiltonian is preserved. That is, \( \tilde{H} \) and the original Hamiltonian \( H \) are unitarily equivalent and thus share the whole energy spectrum, i.e., all the ground- and excited-state energies.

Orbital rotations using general unitary matrices result in complex-valued Hamiltonians and wavefunctions. However, by restricting to real orthogonal matrices, one can retain a real-valued Hamiltonian and corresponding real-valued wavefunctions. This flexibility enables controlled benchmarking of quantum algorithms under both real and complex-valued settings, depending on the specific application context in quantum chemistry.

\section{\label{sec:Similar}Comparison to Molecular Hamiltonians}
The Hamiltonian we have constructed consists of $O(N^4)$ terms, which is the same as molecular Hamiltonians. In this section, we further study the similarity and the difference to the molecular Hamiltonian from various perspectives. In the following sections, we set $\mu = U/2$ for the orbital-rotated 1D Fermi-Hubbard (ORFH) Hamiltonians. By choosing this particular value for $\mu$, the ground state of this system is known to be in a half-filled state~\cite{essler2005one} which has a number of electrons equal to the total number of sites. Each site can accommodate two electrons, so half of the maximum allowable number of electrons are filled. The details of the setting of all numerical simulations in this section are described in Appendix~\ref{app:details}.

\subsection{Properties of the Hamiltonian operator}
\begin{figure}[htbp]
 \includegraphics[width=.5\textwidth]{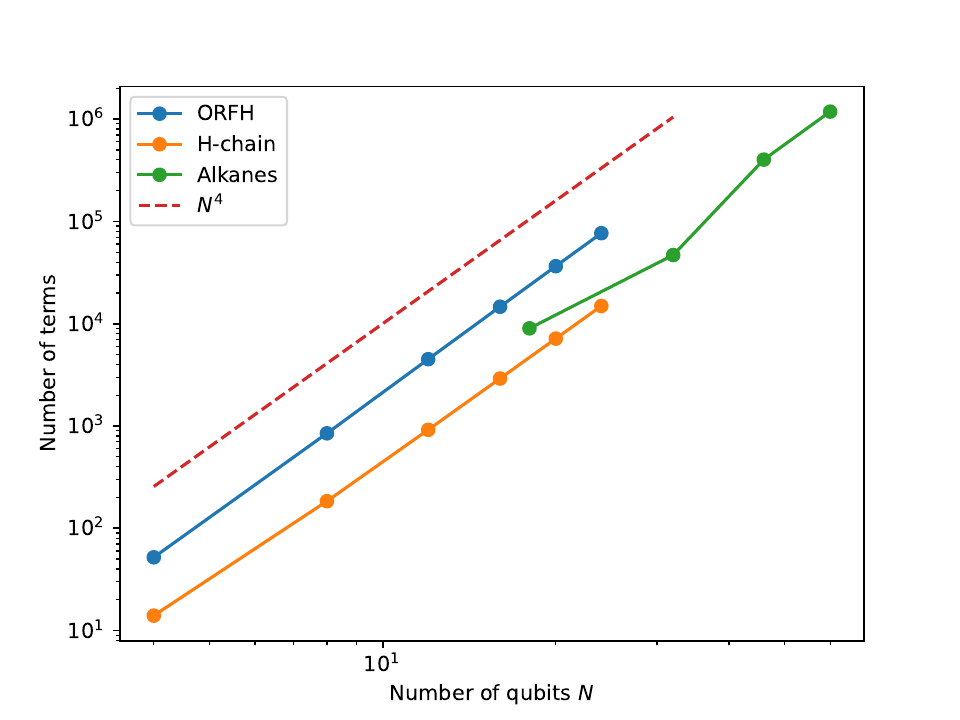}
\caption{Comparison of the number of Hamiltonian terms for the ORFH model and several molecular systems, showing $O(N^4)$ scaling in both cases, but a higher constant prefactor for ORFH. For molecular systems, ``H-chain'' corresponds to linear hydrogen chains \ce{H2}, \ce{H4},...,\ce{H12}, while ``Alkanes'' corresponds to \ce{CH4}, \ce{C2H6}, \ce{C3H8}, and \ce{C4H10}. Dashed line shows the $O(N^4)$ scaling for reference.}
\label{fig:terms}
\end{figure}

We begin by examining the structural characteristics of the orbital-rotated 1D Fermi-Hubbard (ORFH) Hamiltonian in comparison with standard molecular Hamiltonians. We begin by examining the number of terms in the Hamiltonian in more details.
Although both systems exhibit $O(N^4)$ scaling, molecular Hamiltonians tend to have fewer terms than the ORFH Hamiltonian at fixed system size, as shown in Fig.~\ref{fig:terms}. This is because molecular Hamiltonians respect additional symmetries, such as spin, electron number, or spatial symmetries, that eliminate many degrees of freedom. In contrast, the ORFH Hamiltonian includes all possible two-body interaction terms after orbital rotation, leading to a higher prefactor. This distinction becomes particularly relevant in simulation settings where the number of Hamiltonian terms directly impacts the circuit cost, such as in Trotterized time evolution.

\begin{figure}[htbp]
\begin{minipage}{0.5\textwidth}
\subfloat[][1-norm]{
\includegraphics[width=\textwidth]{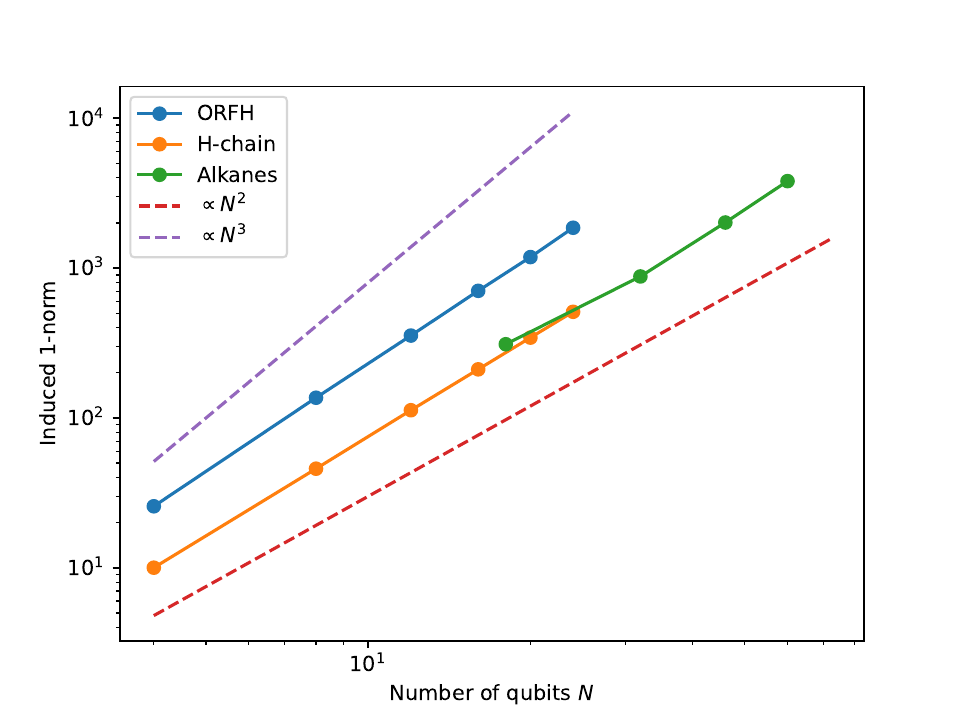}}
\end{minipage}\\
\begin{minipage}{0.5\textwidth}
\subfloat[][2-norm]{
\includegraphics[width=\textwidth]{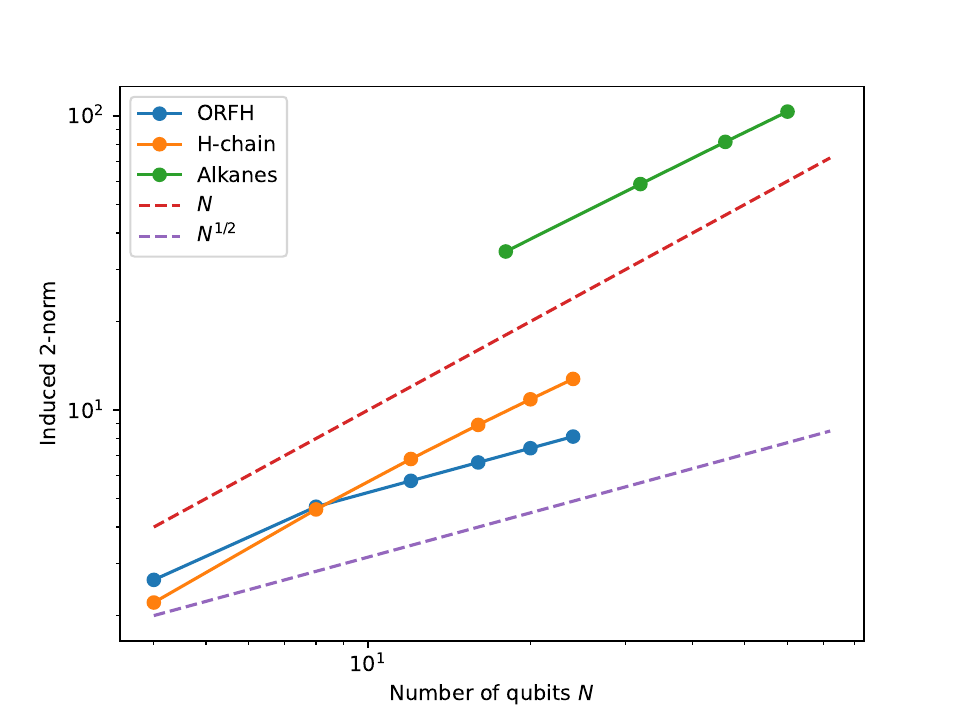}}
\end{minipage}
\caption{Comparison of the induced 1-norm (a) and 2-norm (b) of the ORFH Hamiltonian and molecular Hamiltonians. ``H-chain'' corresponds to linear hydrogen chains \ce{H2}, \ce{H4},...,\ce{H12}, while ``Alkanes'' corresponds to \ce{CH4}, \ce{C2H6}, \ce{C3H8}, and \ce{C4H10}. Dashed lines indicate scalings for reference.}
\label{fig:norm}
\end{figure}

We also analyze the norm of each Hamiltonian to compare their spectral characteristics. Here, we fix $t=1$ and $U=1$ in the ORFH model. While the optimal $U/t$ should ideally be chosen based on the correlation energy (as discussed in Section~\ref{sec:Similar}), we defer that discussion why we choose $U/t=1$ and focus here on the norms. 
Here, we employ the induced $p$-norm of the fermionic Hamiltonian $H$ (without any fermion-to-qubit mapping) with $p=1,2$, which is defined as
\begin{align}
    \left(\sum_i c_i^p\right)^{1/p}, && \text{for }H=\sum_i c_i O_i,
\end{align}
where $O_i$ are product of creation and annihilation operators and $c_i$ are coefficients.
Figure~\ref{fig:norm} shows the induced 1-norm and 2-norm of the Hamiltonian operators. The 1-norm exhibits similar scaling between ORFH and molecular Hamiltonians, though its absolute value tends to be larger in the ORFH case. In contrast, the 2-norm shows noticeably different scaling and smaller magnitudes for ORFH. It is notable, though, that the magnitude of the 2-norm is different even among the molecular Hamiltonians of the hydrogen chains and the alkanes.

These differences in 1- and 2-norms between ORFH and molecular Hamiltonians suggest that they have different distributions of operator coefficients, making it challenging to match both 1-norm and 2-norm simultaneously through parameter tuning of the ORFH Hamiltonian. For the remainder of this work, we therefore simply set $t=1$ as a fixed energy scale and discuss the choice of $U/t$ in the following sections.

\subsection{Hartree-Fock Calculation and Electronic Correlation}

We apply Hartree-Fock (HF) calculations to the ORFH Hamiltonian and analyze its electronic correlation properties, comparing them to those of hydrogen chain (H-chain) molecules.

The HF method provides a mean-field approximation of the electronic ground state, where each electron moves in an average field generated by all others. While computationally efficient, this method neglects the dynamic correlation between electrons. The resulting HF energy $E_{\mathrm{HF}}$ thus overestimates the true ground-state energy $E_{\mathrm{exact}}$, and their difference is referred to as the electronic correlation energy:
\begin{equation}
E_{\mathrm{corr}} = E_{\mathrm{exact}} - E_{\mathrm{HF}}.
\end{equation}
Note that $E_{\mathrm{corr}}$ depends on the chosen basis and does not capture all forms of correlation, some of which are implicitly included in\footnote{As we will discuss shortly, the ORFH Hamiltonian has no explicit spin symmetry, so the HF calculation corresponds to so-called the ``Unrestricted'' Hartree-Fock (UHF) calculation. UHF captures some of the electronic correlations, particularly the one called static correlation.} $E_{\mathrm{HF}}$.

To study how $E_{\mathrm{corr}}$ behaves in ORFH model, in comparison with other systems such as the H-chain Hamiltonian, HF calculations are performed on both systems at half-filling. Since spin is not explicitly conserved due to orbital mixing in the ORFH Hamiltonian, we employ unrestricted HF in both cases.

\begin{figure}[htbp]
\begin{minipage}{0.5\textwidth}
\subfloat[][Hydrogen chain]{
\includegraphics[width=\textwidth]{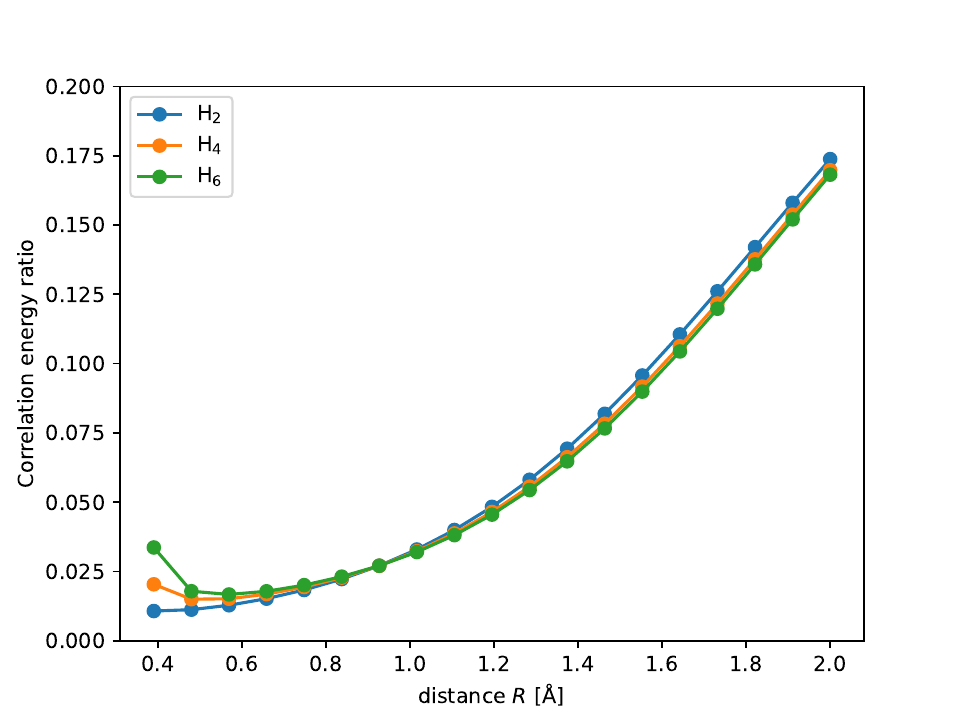}}
\end{minipage}\\
\begin{minipage}{0.5\textwidth}
\subfloat[][Orbital-rotated 1D Fermi-Hubbard model]{
\includegraphics[width=\textwidth]{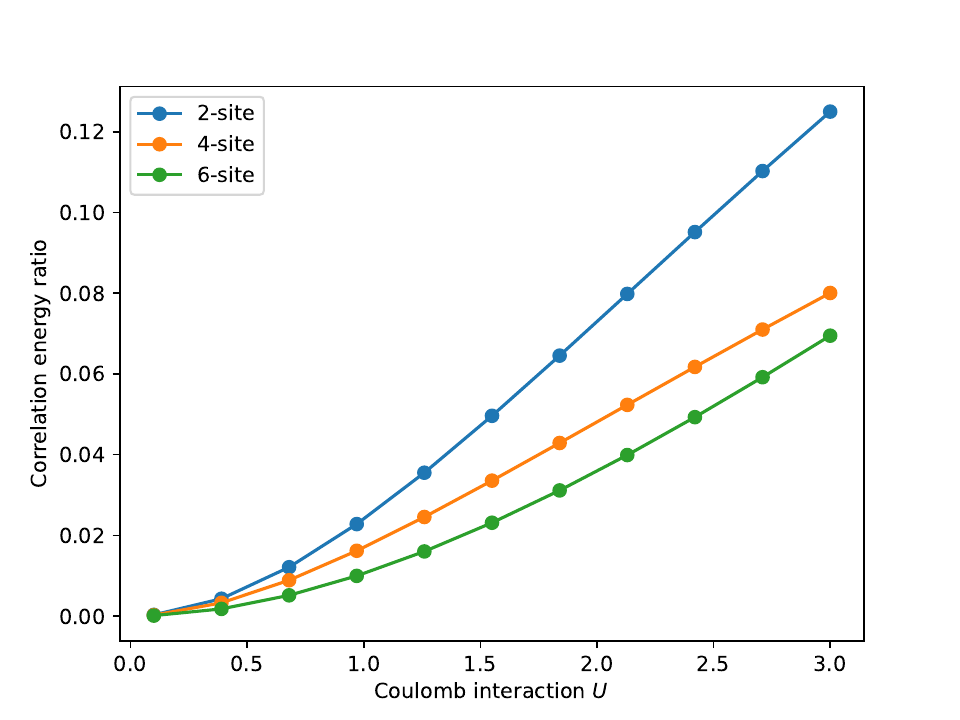}}
\end{minipage}
\caption{(a) Correlation energy ratio $-E_\mathrm{corr}/E_\mathrm{exact}$ of \ce{H2}, \ce{H4}, and \ce{H6} as a function of interatomic distance $R$. (b) Correlation energy ratio $-E_\mathrm{corr}/E_\mathrm{exact}$ of the orbital-rotated FH model with the number of sites $N = 2, 4, 6$ as a function of the Coulomb interaction $U$.}
\label{fig:correlation_energy}
\end{figure}

Figure~\ref{fig:correlation_energy} shows the electronic correlation energy as a function of the atomic spacing $R$ for H-chains and as a function of the Coulomb interaction $U$ for the ORFH model. In both cases, the correlation energy increases monotonically with increasing $R$ or $U$. This trend can be qualitatively understood by the correspondence between the H-chain and the 1D FH model~\cite{sawaya2021constructing}. When $R$ is large, electron hopping becomes negligible and electrons tend to localize on individual atoms—analogous to the large $U$ regime in the FH model. Similarly, smaller $R$ corresponds to increased delocalization, akin to smaller $U$, except for very small $R$, where the simple correspondence appears to break down.

To simulate the regime relevant to typical molecules, we focus on the region where the correlation energy remains moderate. Based on this criterion, we set $U=1$ for the FH model in the remainder of this study. At this value, the electronic correlation is non-negligible yet not too strong, aligning with the correlation strength often encountered near equilibrium geometries in molecular systems.

\subsection{\label{sec:grouping}Grouping of Pauli Terms}

In near-term quantum algorithms such as the variational quantum eigensolver (VQE), reducing the number of measurements is essential for achieving accurate results within limited quantum resources. One widely used strategy is grouping Pauli terms in the Hamiltonian into sets that can be measured simultaneously. The ORFH Hamiltonian contains $O(N^4)$ Pauli terms after the Jordan-Wigner transformation, similar in scaling to molecular electronic Hamiltonians. Accordingly, we expect the performance of grouping methods to exhibit similar trends. We compare three grouping strategies: qubitwise commuting~\cite{mcclean2016theory}, general commuting~\cite{gokhale2019minimizing} and basis rotation grouping~\cite{huggins2021efficient}.

To quantify the performance, we estimate the number of measurement shots required to achieve energy accuracy $\epsilon=0.001$ (in the unit of Hartree for hydrogen chains while the energy for ORFH Hamiltonians have arbitrary unit) in the ground-state energy estimate. The required number of shots $M$ to achieve accuracy $\epsilon$ is given by
\begin{equation}
\label{eq:shot_estimation}
M = \frac{K}{\epsilon^2},
\end{equation}
where the prefactor $K$ depends on the grouping and shot allocation scheme. For optimally allocated shots across groups, $K$ is expressed as
\begin{equation}
\label{eq:K}
K = \left[\sum_G\sqrt{\sum_{\alpha, \beta \in G} h_\alpha h_\beta \mathrm{Covar}(P_\alpha, P_\beta)}\right]^2,
\end{equation}
where $P_\alpha$ and $P_\beta$ are Pauli strings in group $G$, and $\mathrm{Covar}(P_\alpha, P_\beta)$ denotes their covariance~\cite{rubin2018application, gonthier2022measurements}.

\begin{figure}[htbp]
    \includegraphics[width=0.5\textwidth]{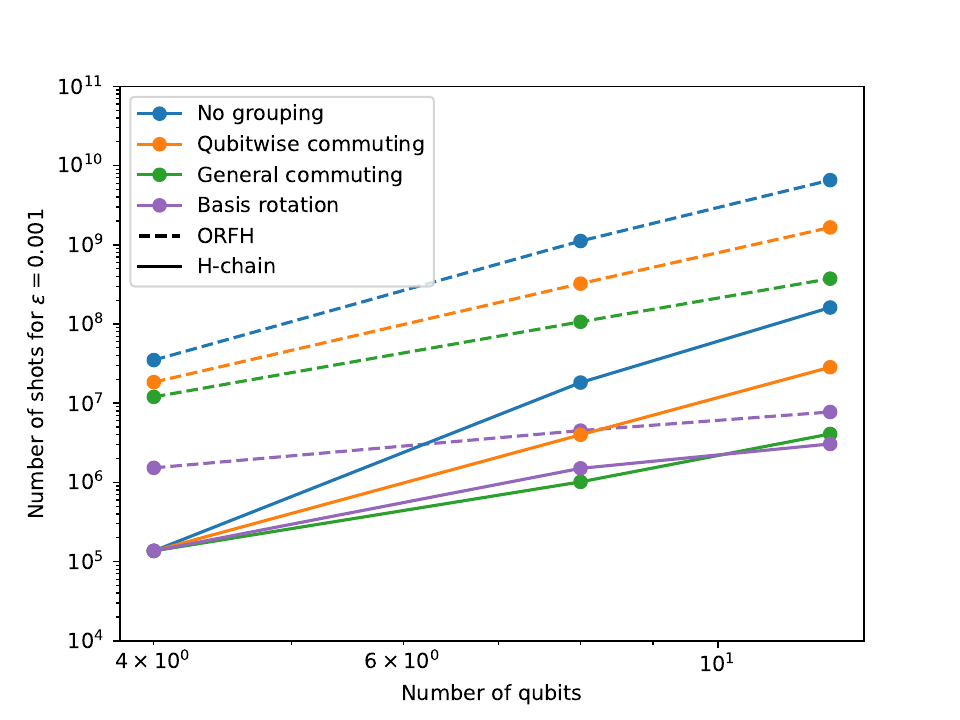}
  \label{fig:shots}
\caption{
Comparison of grouping methods applied to the hydrogen chains (H-chain) and the orbital-rotated 1D Fermi-Hubbard (ORFH) model, up to 12 qubits. The number of shots required to achieve energy accuracy $\epsilon=0.001$ (in the unit of Hartree for H-chain and arbitrary unit for ORFH) is shown. Each plot corresponds to a combination of a model (line style) and a grouping method (color).
}
\label{fig:grouping}
\end{figure}

Figure~\ref{fig:grouping} shows the required number of shots up to 16 qubits, using exact ground states to compute variances and covariances. The overall trends, including the relative ranking of grouping strategies except for the basis rotation grouping, are consistent between the H-chain and the ORFH model. This supports the utility of the ORFH model as a molecular surrogate for benchmarking grouping and measurement strategies to some extent.

A notable difference would be the basis rotation grouping, which shows markedly improved performance for the ORFH model. This may be because the basis rotation step partially reverses the orbital rotation used in constructing the Hamiltonian. In that case, using the ORFH Hamiltonian as a benchmark may introduce a bias that favors the basis rotation method, highlighting a potential limitation of this model for evaluating grouping strategies.

\section{\label{sec:Numarical}Benchmarking Ground-State Solvers}

In this section, we benchmark two widely used algorithms for computing ground states of quantum systems: the variational quantum eigensolver (VQE)~\cite{peruzzo2014variational}, which is a conventional algorithm for near-term quantum devices, and the Density Matrix Renormalization Group (DMRG)~\cite{white1992density, white1993density}, a classical method that is highly efficient for one-dimensional systems. Both methods are applied to the orbital-rotated 1D Fermi-Hubbard (ORFH) Hamiltonian. For comparison, we also apply VQE to H-chain Hamiltonians, and DMRG to the original (unrotated) 1D Fermi-Hubbard model. This allows us to assess the extent to which orbital rotation affects algorithmic difficulty and performance.

\subsection{Variational Quantum Eigensolver}

We first apply VQE to the H-chain Hamiltonians and the ORFH Hamiltonians. VQE is a hybrid quantum-classical algorithm that approximates the ground state of a Hamiltonian $H$ using a parameterized quantum state (ansatz) $|\psi(\vec{\theta})\rangle$. The parameters $\vec{\theta}$ are optimized to minimize the expectation value of the Hamiltonian:
\begin{equation}
E(\vec{\theta}) = \frac{\langle \psi(\vec{\theta})|H|\psi(\vec{\theta})\rangle}{\langle \psi(\vec{\theta})|\psi(\vec{\theta})\rangle} \geq E_0,
\end{equation}
where $E_0$ is the exact ground-state energy. Optimization is performed by a classical routine using measurement outcomes from the quantum computer.

\begin{figure}[htbp]
  \begin{minipage}{0.5\textwidth}
       \subfloat[][Linear \ce{H4}]{
         \includegraphics[width=\textwidth]{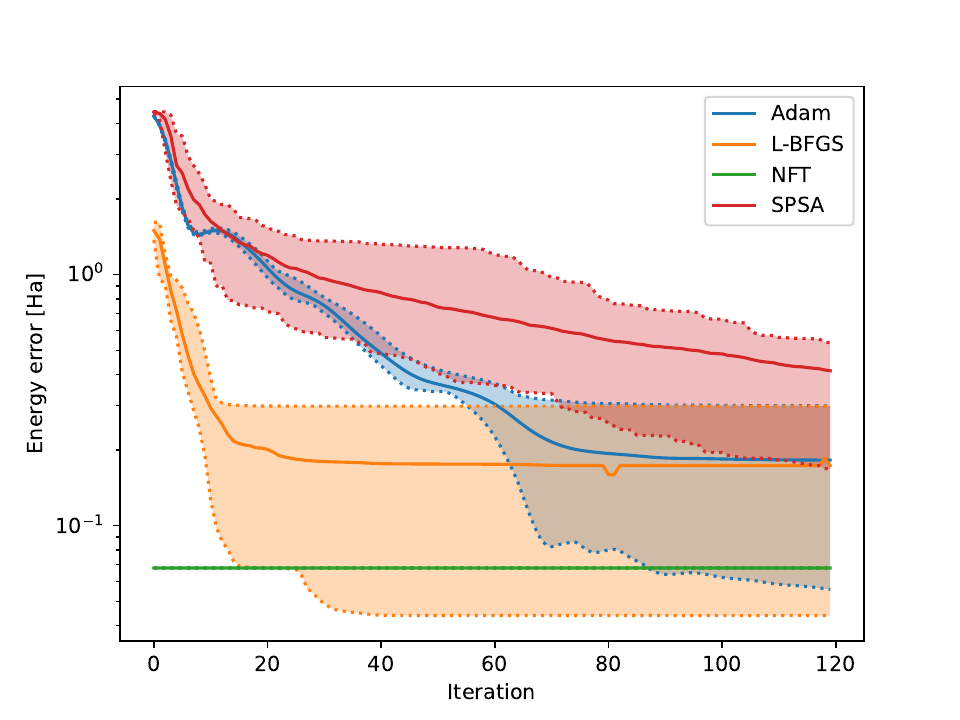}}
  \end{minipage}\\
  \begin{minipage}{0.5\textwidth}
       \subfloat[][Four-site Orbital-rotated 1D Fermi-Hubbard (ORFH)]{
         \includegraphics[width=\textwidth]{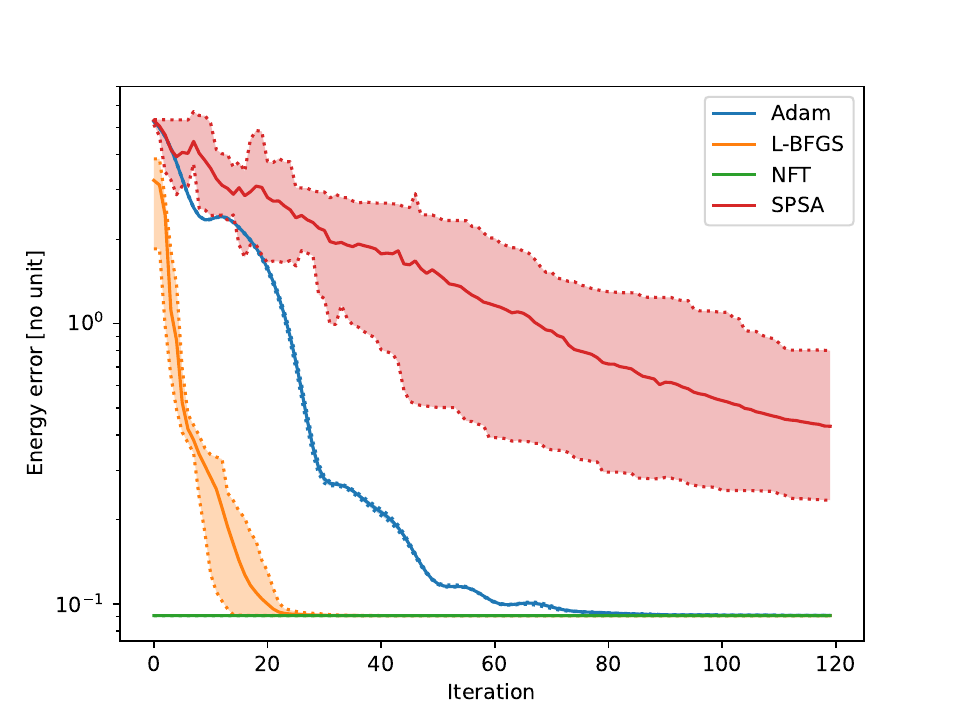}}
  \end{minipage}
 \caption{Results of the VQE using various optimizers applied to (a) the linear \ce{H4} system and (b) the four-site ORFH Hamiltonian with the hardware-efficient ansatz. 
 Dotted lines on the edge of the colored regions show maximum and minimum values at each iteration, and solid lines show mean values.
 Note that NFT optimizes almost perfectly in the first iteration, making the plot almost flat.}
\label{fig:vqe}
\end{figure}

We prepare the H-chain Hamiltonians as in Sec.~\ref{sec:Similar}, and use a hardware-efficient ansatz~\cite{kandala2017hardware} implemented in QURI Parts~\cite{quri_parts}. The expectation values are computed exactly using the Qulacs~\cite{suzuki2021qulacs} statevector simulator as the backend. We compare several optimizers: Adam~\cite{kingma2014adam}, L-BFGS~\cite{liu1989limited}, NFT~\cite{nakanishi2020sequential}, and SPSA~\cite{spall1998implementation}, all implemented via QURI Parts~\cite{quri_parts}. The results for 8-qubit systems (\ce{H4} for the hydrogen chain, or four-site for the ORFH Hamiltonian) are shown in Fig.~\ref{fig:vqe}.
For each optimizer, we executed ten trials, and maximum, minimum, and mean value at each iteration are shown.

From the results, we observe that the convergence behavior is qualitatively similar across optimizers for both Hamiltonians. Although the H-chain exhibits larger variances among ten trials, the relative ranking among optimizers is consistent with that observed in the ORFH case. This suggests that the ORFH model retains key optimization characteristics of molecular systems in the VQE context, and can serve as a useful testbed for comparing optimization strategies.

\subsection{\label{sbsec:dmrg}DMRG}

We now evaluate how the orbital rotation impacts classical ground-state solvers by comparing the performance of DMRG on the original 1D FH Hamiltonian and its orbital-rotated counterpart.

DMRG is a classical variational algorithm based on the matrix product state (MPS) formalism~\cite{white1992density, white1993density, schollwock2011density}, and is widely regarded as the method of choice for 1D strongly correlated systems. We perform finite-system DMRG using the ITensor library~\cite{fishman2022itensor}, on the 16-site systems (both rotated and unrotated) with various maximum bond dimensions of 16 to 256, which is slightly beyond the capability of standard laptops to perform exact diagonalization.

\begin{figure}[htbp]
 \includegraphics[width=.5\textwidth]{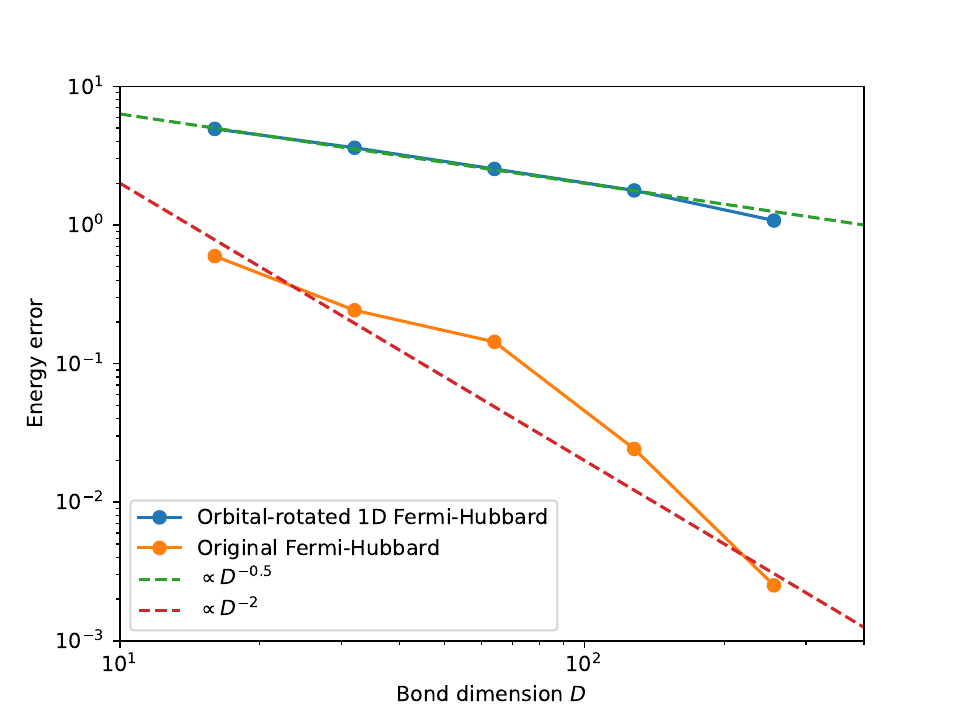}
\caption{Energy errors $E_{\mathrm{error}}=E_{\mathrm{DMRG}}-E_{\mathrm{exact}}$ for the original FH Hamiltonian and the orbital-rotated FH Hamiltonian, obtained by DMRG calculations with MPS bond dimensions of 16, 32, 64, 128 and 256.}
\label{fig:dmrg}
\end{figure}

Figure~\ref{fig:dmrg} shows the relative energy error
$E_{\mathrm{diff}} = E_{\mathrm{DMRG}} - E_{\mathrm{exact}}$
between the DMRG-computed energy and the exact ground-state energy (obtained via Bethe ansatz). The ORFH consistently exhibits significantly larger energy errors, often by orders of magnitude, than the original FH across all bond dimensions studied. Furthermore, the original FH shows more systematic and rapid improvement in accuracy as the bond dimension increases. These results indicate that the orbital rotation increases the entanglement complexity of the ground state, making the ORFH a more challenging problem for classical algorithms such as DMRG as well as for quantum algorithms. From a theoretical perspective, it would be valuable to further investigate the scaling behavior and the underlying principles for this difference, particularly given that exact solutions for both Hamiltonians are available via the Bethe ansatz.

\section{\label{sec:Concl}Conclusion}

In this work, we proposed the orbital-rotated one-dimensional Fermi-Hubbard (ORFH) Hamiltonian as a benchmarking problem for quantum chemistry algorithms. By applying an orbital rotation to the exactly solvable one-dimensional Fermi-Hubbard model, we constructed a Hamiltonian that preserves the exact ground-state energy while increasing the number of terms from $O(N)$ to $O(N^4)$, thereby matching the scaling observed in molecular Hamiltonians. This enables performance evaluation of quantum chemistry algorithms on large, exactly solvable systems with realistic operator structure for both quantum and classical computing.

We investigated the similarity and differences between the ORFH and molecular Hamiltonians from multiple perspectives. In particular, we examined the norm of the Hamiltonian operator and the electronic correlation energy to set the parameters for the ORFH Hamiltonian.
Benchmarking grouping methods and optimizers for the variational quantum eigensolver (VQE) reveals the qualitative match of the results between the ORFH and H-chain Hamiltonians, while we found quantitative discrepancies as well.

Finally, we benchmarked the density matrix renormalization group (DMRG) method on the original and orbital-rotated FH Hamiltonians. Our results show that the orbital rotation substantially increases the difficulty for DMRG, likely due to increased entanglement in the ground state. This further highlights the value of the ORFH Hamiltonian as a nontrivial benchmark not only for quantum algorithms but also for classical methods.

As a direction for future work, it would be interesting to constrain the orbital rotation to better match characteristics of molecular Hamiltonians, such as preserving spin symmetry or aligning the 1- and 2-norms of the Hamiltonian. Such constraints may further improve the utility of the ORFH Hamiltonian as a benchmarking problem.

Another promising direction is to investigate DMRG performance on the ORFH Hamiltonian under varying parameters, such as the interaction ratio $U/t$ or the strength of the orbital rotation. These studies could reveal transitions in the scaling behavior of energy errors with bond dimension $D$, offering valuable theoretical insights, particularly into how entanglement structures differ between molecular systems and the Fermi-Hubbard model.

\section{Acknowledgements}
The authors would like to thank Kosuke Mitarai and Wataru Mizukami for their valuable discussion while formulating the problem. The authors also thank Ming-Zhi Chung, Ryosuke Imai, Shoichiro Tsutsui, Yuichiro Hidaka and Yuya O. Nakagawa for valuable information and comments. This work was performed for Council for Science, Technology and Innovation (CSTI), Cross-ministerial Strategic Innovation Promotion Program (SIP), ``Photonics and Quantum Technology for Society 5.0''(Funding agency : QST).

\appendix

\section{\label{app:details}Details of numerical simulations}
The orbital-rotated 1D Fermi-Hubbard (ORFH) Hamiltonian in this paper was constructed as follows. First, we generate the 1D Fermi-Hubbard (FH) Hamiltonian represented by annihilation and creation operators using OpenFermion~\cite{mcclean2020openfermion}. Throughout this paper, all ORFH Hamiltonians are constructed by applying the orbital rotation using unitary matrices randomly generated using SciPy~\cite{2020SciPy-NMeth}. For all ORFH Hamiltonians in this paper, unless otherwise stated, we set $t=1$ and $U=1$ so that the ORFH Hamiltonian has similar norm and correlation energy of the ground state as the molecular Hamiltonians. We set $\mu = U/2 = 1/2$ so that the ground state being a half-filled state, as explained in the main text.

For all molecular Hamiltonians in this paper, we use the second-quantized electronic Hamiltonian using the Born-Oppenheimer approximation with Hartree-Fock orbitals.  These are constructed by PySCF~\cite{sun2018pyscf} with the STO-3G minimal basis set and then converted into qubit Hamiltonians through the Jordan-Wigner mapping when necessary. Molecular geometries used in this paper are shown in Table~\ref{tab: mol_geom}. 
The geometries of alkane molecules were taken from the PubChem database~\cite{kim2023pubchem}.
\begin{table}[htbp]
  \caption{Molecular geometries are denoted as ``($\mathrm{X}$, $(x, y, z))$,'' where the 3D coordinates $x$, $y$, and $z$ of atom $\mathrm{X}$ are written in units of angstroms \AA. \label{tab: mol_geom}}
 \begin{tabular}{c|p{0.35\textwidth}}
 \hline \hline
 Molecule & Geometry  \\ \hline
 H$_n$ & (H, (0, 0, 0)), (H, (0, 0, 1.0)), \dots,  (H, (0, 0, $n \times 1.0$)) \\ \hline
 Methane \ce{CH4} &(C, (0, 0, 0)), (H, (0.5541, 0.7996, 0.4965)), (H, (0.6833, -0.8134, -0.2536)), (H, (-0.7782, -0.3735, 0.6692)), (H, (-0.4593, 0.3874, -0.9121)) \\ \hline
Ethane \ce{C2H6} &(C, (-0.756, 0, 0)), (C, (0.756, 0, 0)), (H, (-1.1404, 0.6586, 0.7845)), (H, (-1.1404, 0.3501, -0.9626)), (H, (-1.1405, -1.0087, 0.1781)), (H, (1.1404, -0.3501, 0.9626)), (H, (1.1405, 1.0087, -0.1781)), (H, (1.1404, -0.6586, -0.7845)) \\ \hline
Propane \ce{C3H8}&(C, (0, -0.5689, 0)), (C, (-1.2571, 0.2844, 0)), (C, (1.2571, 0.2845, 0)), (H, (0, -1.2183, 0.8824)), (H, (0, -1.2183, -0.8824)), (H, (-1.2969, 0.9244, 0.8873)), (H, (-1.2967, 0.9245, -0.8872)), (H, (-2.1475, -0.352, -0.0001)), (H, (2.1475, -0.352, 0)), (H, (1.2968, 0.9245, 0.8872)), (H, (1.2968, 0.9245, -0.8872)) \\ \hline
Butane \ce{C4H10}&(C, (-0.563, 0.516, 0.0071)), (C, (0.563, -0.5159, 0.0071)), (C, (-1.9293, -0.1506, -0.0071)), (C, (1.9294, 0.1505, -0.0071)), (H, (-0.4724, 1.1666, -0.8706)), (H, (-0.4825, 1.1551, 0.894)), (H, (0.4825, -1.1551, 0.894)), (H, (0.4723, -1.1665, -0.8706)), (H, (-2.0542, -0.771, -0.9003)), (H, (-2.0651, -0.7856, 0.8742)), (H, (-2.7203, 0.606, -0.0058)), (H, (2.0542, 0.7709, -0.9003)), (H, (2.7202, -0.6062, -0.0059)), (H, (2.0652, 0.7854, 0.8743))
\\
\hline \hline
 \end{tabular}
\end{table}
All numerical simulations involving quantum circuit simulations have been conducted by using QURI Parts~\cite{quri_parts}, an open source library for developing quantum algorithms. 


\bibliographystyle{IEEEtran}
\bibliography{orbital_rotated}
\end{document}